# LOW TEMPERATURE MAGNETIC AND DIELECTRIC PROPERTIES CORRELATION IN FE- DOPED COPPER (II) OXIDE CERAMICS FOR POTENTIAL DEVICE APPLICATION.

Kumar Brajesh[1], Ritamay Bhunia[2], Shashikant Gupta,[2], Rajeev Gupta[3], Ambesh Dixit[4] and Ashish Garg[1]

[1]Department of Sustainable Energy Engineering
[2]Department of Material Science Engineering
[3]Department of Physics
Indian Institute of Technology Kanpur, Kanpur-208016 (India)
[4]Department of Physics, Indian Institute of Technology Jodhpur, Karwad – 342037 (India)

**Abstract**

The bulk samples of CuO and Fe-doped CuO were synthesized by ceramics methods. Structural and compositional analyses were performed by using X-ray diffraction, SEM, and EDAX. Through this manuscript, we are going to report the effect of trivalent iron doping ($Fe^{3+}$) in copper (II) oxide ($Cu_{0.95}Fe_{0.05}O$) bulk samples on magnetic and dielectric behavior. The paramagnetic phase has been established in CuO as a result of $Fe^{3+}$ doping. The strong correlation between magnetic and dielectric properties indicated spin-polaron interaction at the transition temperature. Bulk CuO and also $Cu_{0.95}Fe_{0.05}O$ exhibit the multiferroic phase in a narrow temperature range (190 K to 230 K). Two transitions happened from a paramagnetic-paraelectric phase to incommensurate or asymmetrical antiferromagnetic (AF) and ferroelectric state near highest Neel temperature ($T_{N1}$) ~230 K and another second phase transition, the order of AF phase transformed to commensurate AF phase and ferroelectricity disappeared at around the Neel temperature ($T_{N2}$) ~210 K in all samples. This $Cu_{0.95}Fe_{0.05}O$ would show its potential in the spintronic application for a high dielectric constant with low loss and high magnetic susceptibility.

## 1. Introduction

Copper oxide have attracted extensive attention recently, because of their unique properties. Indeed, the CuO has been studied as a p-type semiconductor with a band gap value 1.2 eV, because of the natural abundance, low-cost production processing, nontoxic nature, and its reasonably good electrical and optical properties [1-10]. Because of its interesting properties, the Copper oxide material is an important candidate for a variety of practical applications in the electronic and the optoelectronic devices, such as micro-electromechanical systems, spintronics, solar cells, gas sensors [11] and photo thermal applications. Bulk CuO has been studied for variety of applications in last few years due to its electrical and optical properties and it exhibits the multiferroic phase in a narrow temperature range (210 K to 230 K). A transition happens from a paramagnetic-paraelectric phase to incommensurate or asymmetrical antiferromagnetic (AF) and ferroelectric state near highest Neel temperature ($T_{N2}$) ~230 K. In case of second phase transition, the order of AF phase transforms to commensurate AF phase and ferroelectricity disappears at around the Neel temperature ($T_{N1}$) 210 K.[12,13].

Doping of transition-metal ions in CuO lattice has led towards interesting ferromagnetic behaviours at or above room temperature[14,15]. Researchers demonstrated enhanced magnetization by doping of Fe in CuO due to the high magnetic moment per atom of Fe through various synthesis technique: co-precipitation, hydrothermal, spray-deposition, sol-gel [16-20]. However, they reported about nanoparticles, nanorods and thin films, where the high yield was not possible. There is no such report as per our knowledge in bulk Fe doped CuO. In a recent work, Borzi et al. failed to observe any ferromagnetic behaviour with addition of Fe-dopant in CuO, the corresponding magnetic susceptibility was increased due to the doping of Fe [21]. Whereas, Li et al. was able to observe room temperature ferromagnetism in Fe doped CuO. Their magnetic hysteresis loops and magnetization-temperature studies pointed out that their materials were mictomagnetic in ferromagnetic domains due to the ferromagnetic coupling between the Fe dopant in $Cu_{1-x}Fe_xO$ nanocrystals synthesized by combustion method.[22] In another work, lower temperature ferromagnetism was reported in $Cu_{0.95}Fe_{0.05}O$ nanocrystals prepared by solid-state reaction, where the origin of ferromagnetism was the super-exchange interaction between two metal ions[23]. Meneses et al. tried to dope transition metal ($Ni^{2+}$ and $Fe^{3+}$) in CuO matrix by coprecipitation method. They demonstrated the suppression of AF ordering in case of Fe doping through their temperature dependent magnetization (M-T) studies[24]. However, there is no unique conclusion behind the origin of magnetization for doped CuO in case of bulk samples and nano-scale samples.

In this article, we have doped with Fe (5%) in paramagnetic CuO through solid-state reaction and thoroughly characterized its structural, chemical, compositional, dielectric and magnetic behaviors. We critically analysis the temperature-dependent magnetic susceptibility and dielectric behavior and correlated them through the corresponding transition temperature.

**2. Experimental**

Solid-state reaction synthesis has been followed to prepare polycrystalline CuO, $Cu_{0.95}Fe_{0.05}O$, and $Cu_{0.995}Fe_{0.005}O$ powder samples. The powders of high purity Iron (III) oxide (99.99%), Copper(II) oxide (99.999%) (all from Sigma Aldrich) are mixed in a stoichiometric ratio using a mortar and pestle. Homogenized powder mixtures are pelletized and calcined at 800°C for 8 hours, followed by regrinding, palletization and sintering at 900°C for 10 hours in air. The phase identification of the samples has been carried out using PANanalytical's X'Pert PRO x-ray diffractometer. The Nova Nano SEM field-emission scanning electron microscope (FE-SEM) is used for microstructural analysis of the samples. Energy dispersive X-ray analysis (EDAX) and electron probe microanalysis (EPMA) in SEM are used for elemental mapping of the samples. The impedance measurements are carried out using an Agilent 4294A Precision Impedance Analyzer, where an AC signal is applied across the pellets, and the output is recorded over the frequency domain. For magnetic behavior, measurements is carried out from cryogenic squid [Give Machine name]. X-ray photoelectron spectroscopy (XPS) measurements were performed to analyze the chemical states of the samples using PHI 5000 (Versa Probe II, FEI Inc) and XPSPEAK 4.3 software was used to analyze the spectra. Fourier-transform infrared spectroscopy (FTIR, PerkinElmer) was used to study the functional groups of the samples in range of wavenumbers 400-4000 $cm^{-1}$. Rietveld refinement was carried out using FULLPROF software [25].

**3. Results and Discussions**

Fig. 1a shows room temperature (RT) XRD patterns for the CuO powder. X-ray diffraction measurements have been carried out to study the crystal structure and crystalline quality. All the diffraction peaks are indexed to crystalline monocline structure with space group C2/c. No other peaks are observed belonging to any impurities such as $Cu(OH)_2$ and $Cu_2O$, indicating the high purity of CuO powders (JCPDS NO-80-1268). Fig. 1b also shows the XRD patterns at room temperature (RT) of $Cu_{0.95}Fe_{0.5}O$ powder. This X-ray diffraction pattern indicates that there is no second impurity phase in the solid powder of $Cu_{0.95}Fe_{0.05}O$.

### 3.1. Structural and Microstructural Analysis at Room Temperature

Figure 1a shows RT XRD patterns for the CuO powder. X-ray diffraction measurements have been carried out to study the crystal structure and crystalline quality. All the diffraction peaks are indexed to crystalline monocline structure with space group C2/c. No other peaks are observed belonging to any impurities such as $Cu(OH)_2$ and $Cu_2O$, indicating the high purity of CuO powders (JCPDS NO-80-1268). Fig. 1b also shows the XRD patterns at RT of $Cu_{0.95}Fe_{0.5}O$ powder. This X-ray diffraction pattern indicates that there is no second impurity phase in the solid powder of $Cu_{0.95}Fe_{0.05}O$.

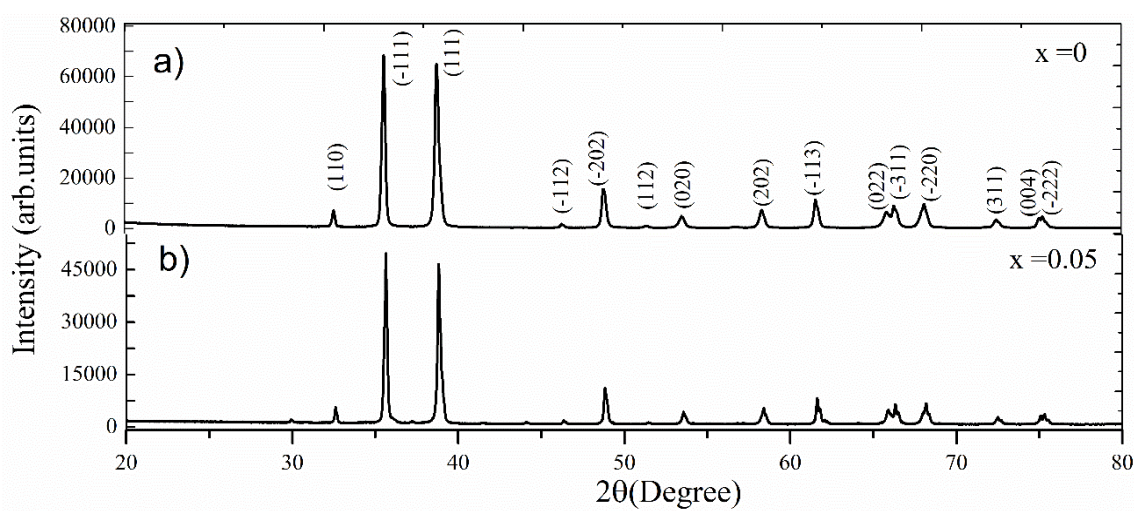

**Figure 1.** (a) and (b) XRD pattern of CuO and $Cu_{0.95}Fe_{0.5}O$ at room temperature

We have also done Rietveld refinement of these two samples to analyze the structural properties and to determine the crystal parameters. For these refinements, we have used XRD diffraction data in the 2θ range 20-80° and monoclinic structure with space group C2/c. The Rietveld refinement profile of the XRD of x=0 and x=0.05 are shown in Fig.2(a) and 2(b). The crystal parameters for these two samples have been presented in Table 1.

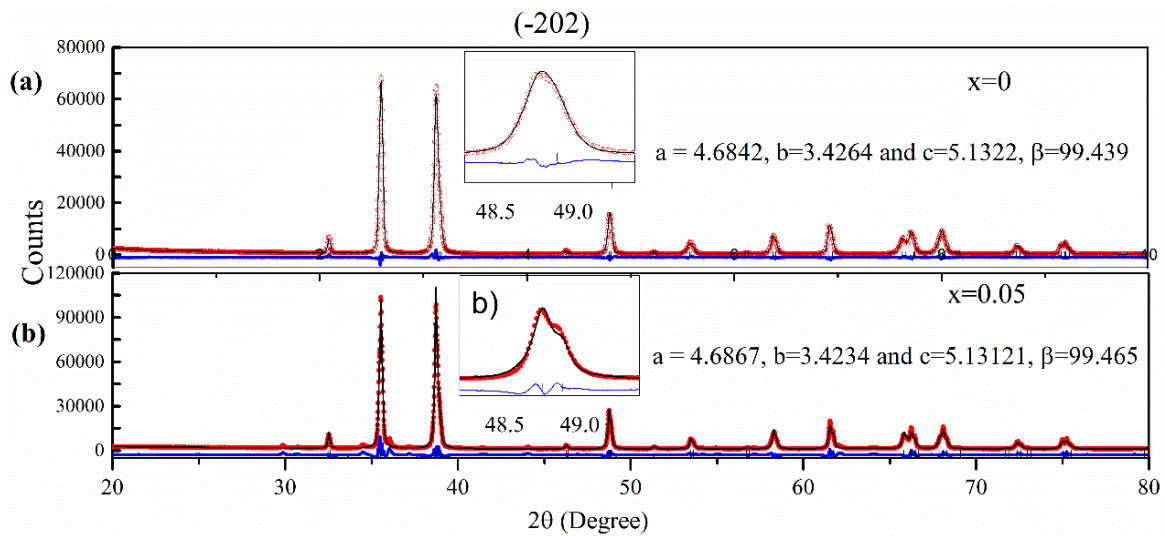

**Figure 2.** Reitveld refinement of X-ray diffraction patterns for (a)CuO and (b) $Cu_{0.95}Fe_{0.05}O$ ceramics at room temperature.

### 3.2. Chemical state analysis

The elemental composition and oxidation states of the Pristine (CuO) and Fe doped CuO samples were analyzed using XPS. Figure 3(a-e) shows high-resolution core level lines of the Cu 2p, Fe2p, and O1s, respectively, for both the samples. Fig 3(a) shows that the high–resolution Cu2p core level splits into two main peaks as $Cu2p_{3/2}$ and $Cu2p_{1/2}$ peaks, at a binding energy of 933.55eV and 953.65 eV, respectively, along with two satellite peaks are observed at higher binding energy at 941.50eV and 961.75eV, respectively. The satellite peaks are characteristic of $Cu^{2+}$ valance state in copper halides [26] and confirm that $Cu^{2+}$ valance state in this sample. Now, in the doped sample(x=0.05), the main peaks are observed at 933.47eV and 953.53eV, whereas satellite peaks are observed at 941.24eV and 961.67eV, respectively shown in Figure 3(c). Interestingly, we have observed that the Cu2p peaks for Fe doped CuO shifted the lower binding energy in comparison to the pristine sample. It may be due to the lower Pauling electronegativity of Fe (1.82eV) as compared to Cu (1.90eV). Therefore, electron transfer from Fe to CuO resultant may shift the peaks to the lower binding energy.

Figure 3b shows the deconvoluted high-resolution XPS peak of O1s for CuO sample at 529.37 eV corresponding to lattice oxygen ($O_L$) and peak at 530.78 eV assigned to surface oxygen vacancies/defects or surface adsorbed oxygen ($O_V$). The relative concentration of oxygen vacancy within both the samples was evaluated from the relative area of $O_v$ peak [$O_v/(O_v + O_L)$] in Figure 3d.

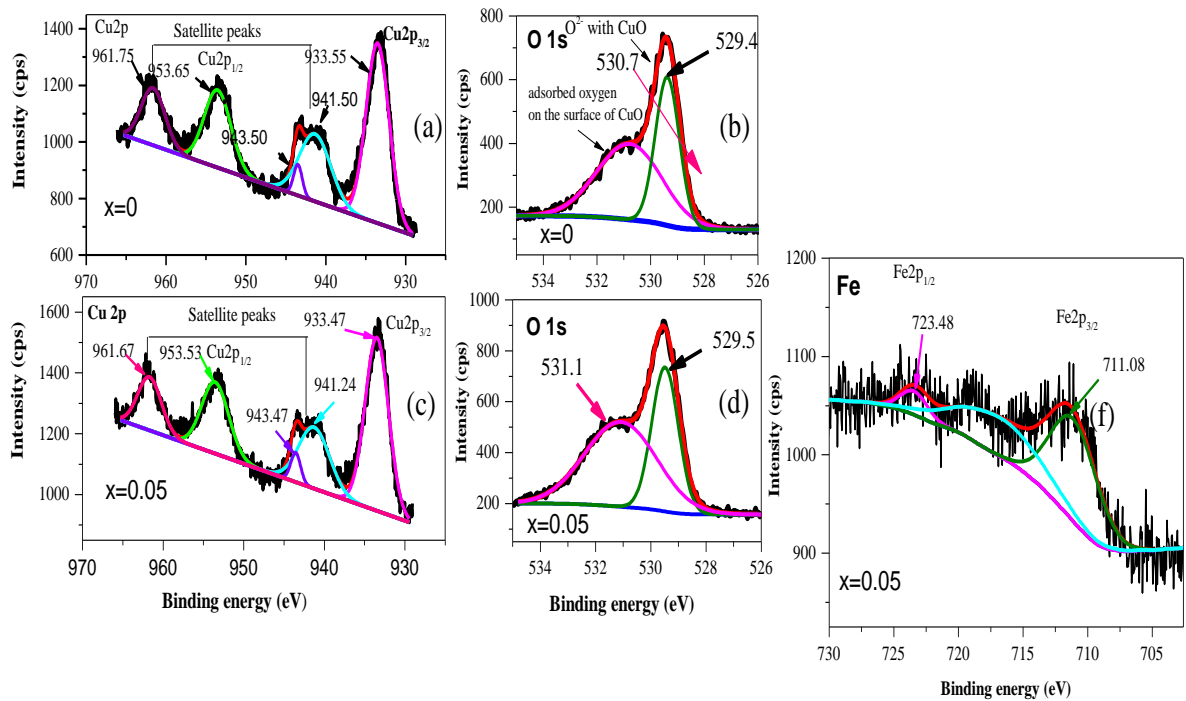

**Figure 3.** High resolution XPS spectra of (a) Cu 2p and (b) O 1s for CuO. High resolution XPS spectra of (c) Cu 2p (d) O 1s and (e) Fe 2p for $Cu_{0.95}Fe_{0.05}O$ ceramic.

Upon evaluation it is observed that oxygen vacancy concentration increases from 58.03 % for CuO samples to 61.04 % for the doped samples, which signifies enhancement in oxygen related defects in doped samples. Figure 3(e) represent the Fe 2p core level also splits in two peaks at ~723.48eV and ~711.08 eV corresponding to $Fe2p3_{/2}$ and $Fe2p1_{/2}$ respectively. The 12.4 Ev splitting in Fe 2p spectra is due to spin –orbit coupling. In the case of $Fe2p3_{/2}$, the range of binding energies are 709.4-710.3 eV and 710.3-711.4 eV for $Fe^{2+}$ and $Fe^{3+}$ respectively[27,28]. The broadness of the $Fe2p3_{/2}$ peaks reveals that mixed of valance state of majority $Fe^{3+}$ and minority $Fe^{2+}$ present in the sample for x=0.05.

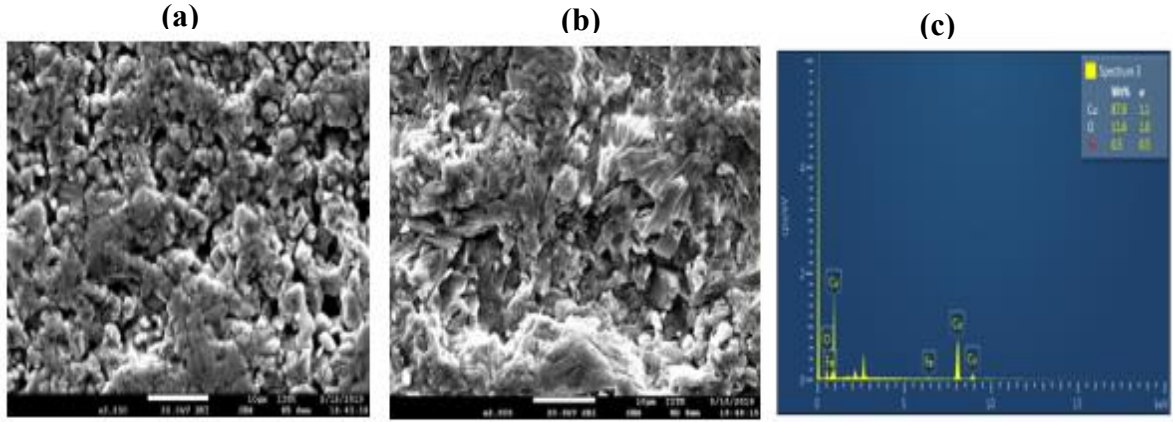

**Figure 4.** SEM images of (a) CuO and (b) $Cu_{0.95}Fe_{0.5}O$ pallet and (c) EDAX spectrum of $Cu_{0.95}Fe_{0.05}O$

Generally, the microstructure has significant influences on the physical properties of condensed matter. The SEM micrographs of CuO and $Cu_{0.95}Fe_{0.05}O$ pallets are depicted in Fig. 4(a) and 4(b), respectively. From these microstructural observations, we can notice that a porous microstructure with small grains and larger pores appears in the grain boundary of CuO. We have observed that some pores and larger grains (~5µm) than CuO in the grain boundary in the case of $Cu_{0.95}Fe_{0.5}O$. EDAX spectrum, shown in Fig. 4(c), not only identifies the elements corresponding to each of its peaks, but the intensity of peaks corresponds element's concentration. The peaks have confirmed the presence of Copper, Oxygen, and iron atoms. The average atomic weight percentage ratio of Cu, O, and Fe in $Cu_{0.95}Fe_{0.5}O$ is 87.9: 11.6: 0.5.

### 3.3. Optical Spectroscopy Studies at room temperature

We have performed the ultraviolet-visible (UV-vis) diffuse reflectance spectroscopy of two samples. Fig. 5a and 5b show the absorption spectra of CuO and Fe doped CuO, respectively. We have estimated the bandgap ($E_g$) from those absorbance spectra using the Kubelka-Munk function[29,30]. We don't observe any significant change in $E_g$ due to iron doping in CuO. However, the bandgap is decreased from 1.3 eV to 1.28 eV in the case of $Cu_{0.95}Fe_{0.05}O$ compared to pure CuO, which is quite apparent due to metal doping.

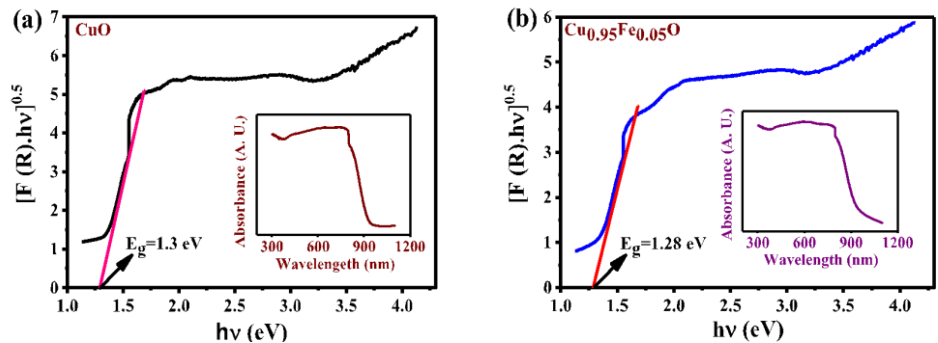

**Figure 5.** Band gap calculation by Kubelka-Munk function of (a) CuO and (b) $Cu_{0.95}Fe_{0.05}O$. Insets show the corresponding optical absorbance spectra.

### 3.4. Magnetic Susceptibility and Transition Temperature

Fig. 6a and 6b show the thermal dependence of the susceptibility ($\chi$=M/H, H= 1000 Oe) in cases of FC and ZFC for CuO and $Cu_{0.95}Fe_{0.05}O$ samples. The graph's nature for bulk CuO is very similar to some previously reported trends for bulk CuO [31-33]. We have determined the Neel temperature ($T_N$) by finding the peaks in ($\partial\chi/\partial T$) vs. T plot. More interestingly, two 3-dimensional phase transition temperatures ($T_N$) has been observed: one at 230 K ($T_{N1}$) and the other at 213 K ($T_{N2}$). $T_{N1}$ and $T_{N2}$ are due to second-order transition with incommensurate ordering and first-order lock-in transition, respectively[37-39]. Below temperature 213 K ($T_{N2}$), the spin structure of CuO is collinear antiferromagnetic. Between temperatures 213 K and 230 K, the spin structure transforms to non-collinear and incommensurate[34-37]. Upon doping of Fe in CuO, we still have found out the two transition temperatures, $T_{N1} \sim 225$ K and $T_{N2} \sim 190$ K, but at lower temperatures than bulk CuO.

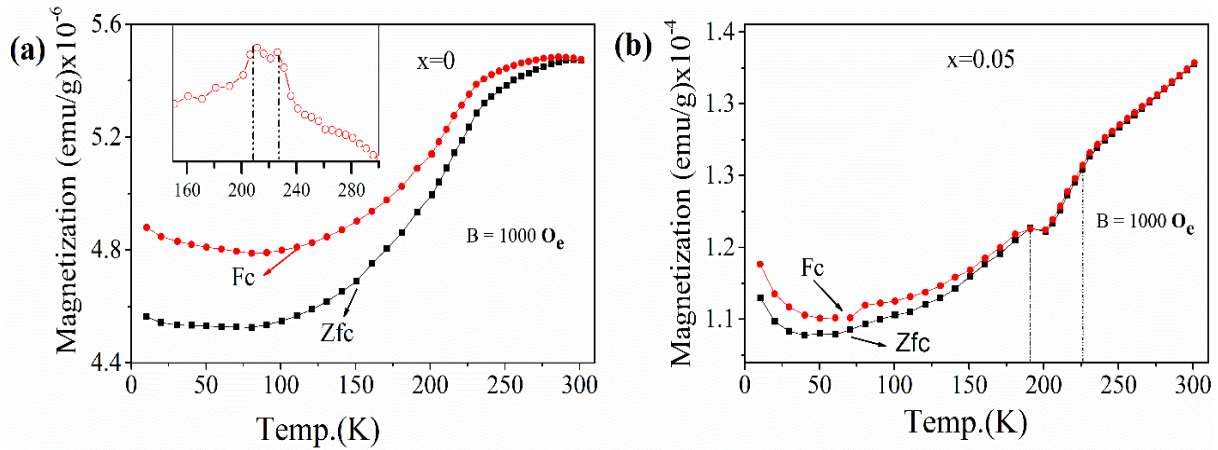

**Figure 6.** Zero field cooling (ZFC) and field cooling at 1000 Oe external field (FC) temperature-dependent magnetic susceptibility of (a) CuO and (b) $Cu_{0.95}Fe_{0.5}O$. Insets indicate the corresponding magnetic transition for CuO.

Observing the two transition temperatures also in $Cu_{0.95}Fe_{0.05}O$ during cooling from room temperature provides proof against our XRD analysis that doping of Fe in CuO did not change the monoclinic structure of CuO. However, considering only magnetic interactions, the transition ordering temperature should not be affected for the substitution of $Cu^{2+}$ by $Fe^{3+}$ ion in the case of $Cu_{0.95}Fe_{0.05}O$. In reality, doping of each Fe atom creates an impurity band, donating an extra electron in the crystal system. These delocalized electrons may mediate an RKKY like interaction among the localized transition metal moments, which may introduce a magnetic frustration and cause a decrease in $T_N$. Trivalent atom ($Fe^{3+}$) doping in CuO also increased the magnetic susceptibility introducing paramagnetic behavior in our $Cu_{0.95}Fe_{0.05}O$ sample. We observed changes in the nature of magnetization curves in cases of FC and ZFC for the $Cu_{0.95}Fe_{0.05}O$ sample also for the paramagnetic contribution due to trivalent atom doping [38].

Figures 7(a) and (b) show that magnetic field-dependent magnetization (M-H) plots at the temp. of 10K and 300K for x=0 and x=0.05. The observed weak ferromagnetism is superimposed by the paramagnetic contribution of the material, and we subtracted the high field paramagnetic contribution, and the corrected hysteresis data is shown in Figure 7a. The M-H loop shows a typical behavior of weak ferromagnetic, where the magnetization does not attain the saturation value even up to the maximum applied field of ~0.6T for both samples.

The coercive field and remnant magnetization decrease with increasing temp for x=0 and x=0.05, respectively, which are shown in the inset of Figures 7(a) and (b). We have also observed that the value of remnant magnetization ($M_r$) and coercive field ($H_c$) is maximum for x=0.05, indicating that stronger exchange interaction happens in the Fe-doped sample.

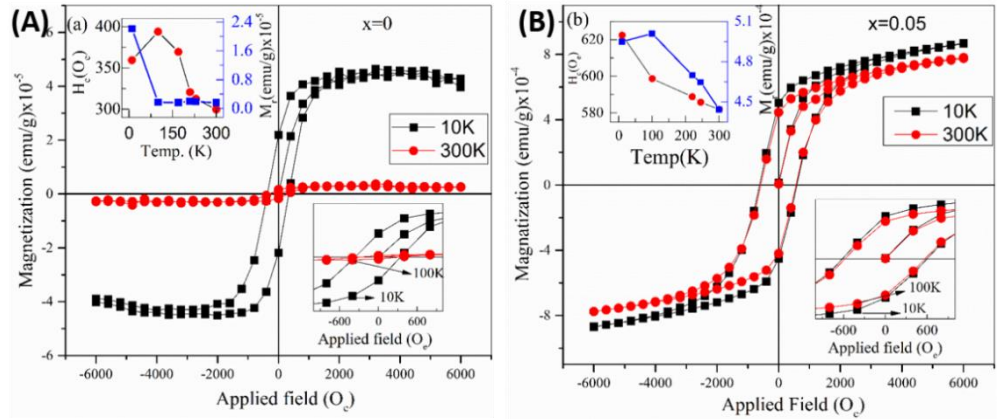

**Figure7.** Magnetic hysteresis curve for (a) CuO and (b) $Cu_{0.95}Fe_{0.05}O$ at the temperature 10K and 300K. Inset fig shows that coercive field and remnant magnetization vs. temp.

The weak ferromagnetism observed in pure CuO is due to the $Cu^{2+} - V_o - Cu^{2+}$ interactions; however, in the case of Fe-doped samples, the $Fe^{2+} - V_o - Fe^{2+}$ and $Fe^{3+} - V_o - Fe^{3+}$ interactions give rise to the strong ferromagnetic behavior in these samples[40]. The enhancement in the ferromagnetic behavior is attributed to the strong $Fe^{3+} - V_o - Fe^{3+}$ interactions that are dominant in our samples as revealed by the XPS measurements. In the case of Fe doped CuO, ferrimagnetic correlation among $Fe^{3+}$ ions substituting the $Cu^{2+}$ ion site may lead to an ordered spin configuration, enhancing the remnant magnetization value and coercive field for x=0.05.

### 3.5. Low Temperature-Dependent Dielectric Properties

The temperature dependence dielectric behavior of pure CuO and Fe doped CuO has been investigated through impedance spectroscopy. The dielectric constant ($\varepsilon'$) can be calculated with the help of the following equation:

$$\varepsilon' = \frac{|Z''| * Thickness\ of\ the\ pellet}{\varepsilon_0 \omega |Z|^2 * Area\ of\ the\ electrode} \tag{1}$$

Where, $Z$ and $Z''$ are the impedance and the complex portion of impedance, respectively, $\omega$ is the angular frequency of given ac signal during measurement and $\varepsilon_0$ is the free space permittivity. The dielectric behavior of any material entirely depends on the polarizability of that particular material. Fig. 8a gives the information about the dielectric behavior of virgin CuO powder for the temperature range 80 K – 300 K. It has been observed that the dielectric constant ($\varepsilon'$) decreased with increasing frequency. Values of both $\varepsilon'$ and $\varepsilon''$ were higher at higher temperatures especially in the lower frequency ranges. The above variation could be ascribed as arising due to possible huge charge accumulation at the electrode-sample interface. The contribution from the charge carriers to the dielectric constant would decrease with increasing frequency due to the high periodic reversal of the field at the interface in the high-frequency region [39].

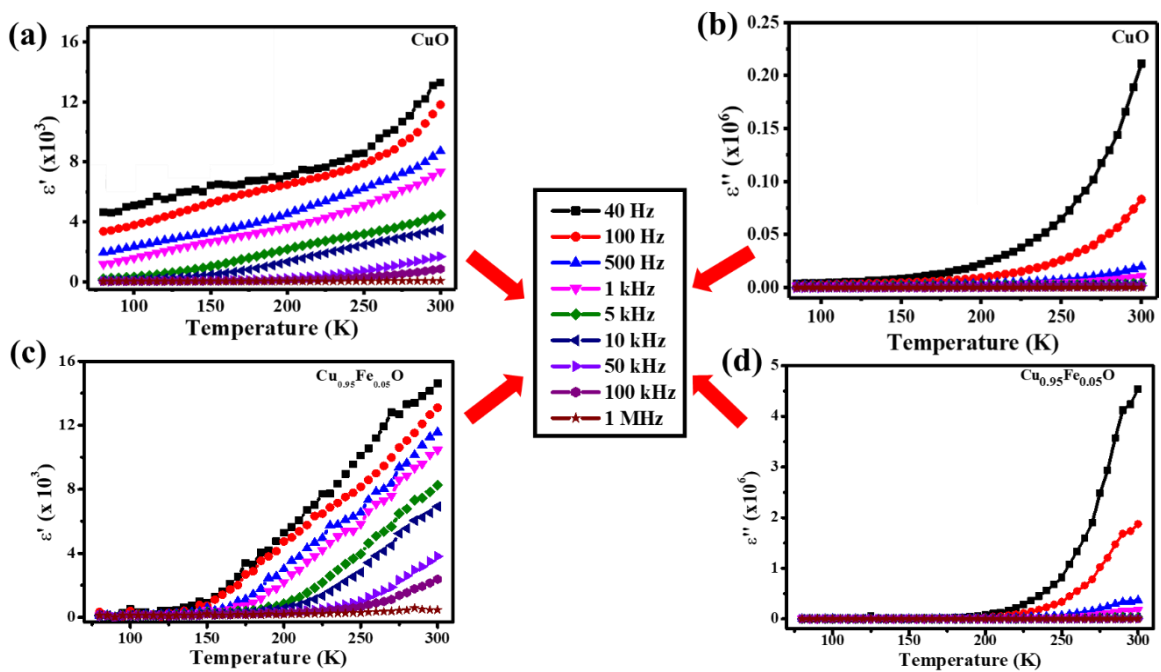

**Figure 8.** (a) Dielectric constant ($\varepsilon'$) and (b) dielectric loss ($\varepsilon''$) with temperature range (80 – 300 K) of CuO and (c) Dielectric constant ($\varepsilon'$) and (d) dielectric loss ($\varepsilon''$) with a temperature range (80 – 300 K) of $Cu_{0.95}Fe_{0.5}O$. Insets shows the corresponding dielectric transition.

The nature of the dielectric curve indicates that the bulk CuO has behaved as dielectric material with significant dielectric loss. This can be attributed to the oxygen vacancies in CuO. Oxygen vacancies may create point defects in crystal, and they are pervasive in any oxides of metals. Oxygen vacancies increase the ac conductivity hence increasing the lossy nature. The dielectric constant of the two samples is nearly the same, but the dielectric loss has decreased during doping of $Fe^{3+}$ in CuO. Generally, in any solid dielectric, there is the probability for four types of polarizations: (1) electronic polarization, (2) space charge or interfacial polarization developed at grain boundaries and electrodes, (3) dipolar polarization between neighboring ion pairs, and (4) atomic polarization due to movement of single ions for development of strain and vacancies[41,42]. Now, above stated second and third types of polarization are strongly temperature-dependent. $Fe^{3+}$ and $Cu^{2+}$ form a dipole-type polarization in the doped sample that further increases the dielectric constant. The dielectric permittivity with temperature because of thermally activated charge carriers' hopping. At low temperature, this thermally activated hopping among polarons and electrons is confined in their respective positions due to low temperature, resulting in low conductivity and low dielectric loss. With an increment of temperature, the lattice vibrations help to activate electron hopping which further leads to polaron hopping[43]. These phenomena increase the dielectric constant and loss factor with stepping up of temperature. The manifestation of the magnetic phase transitions is also seen in our temperature-dependent dielectric observations. The magnetic transition also has been effected the dielectric properties for $Cu_{0.95}Fe_{0.05}O$. After around 190K temperature, the dielectric constant ($\varepsilon'$) increases rapidly in the case of $Cu_{0.95}Fe_{0.05}O$, but the dielectric loss ($\varepsilon''$) does not show any anomaly after that temperature. This is not due to only for thermal effect. We also observed magnetic transition in these materials at this temperature, which suggests a strong spin-polaron coupling in this Fe doped CuO [44]. However, in CuO, the change in dielectric constant is seen after 210 K. Our CuO sample shows the first magnetic transition at 213K. This indicates the magnetically induced polarization in two samples[45]. It is noteworthy to mention that the magnetic effect on dielectric behavior is more in the Fe doped CuO than pristine CuO.

### 3.6. Impedance Spectroscopy Analysis

Temperature-dependent complex impedance spectroscopy helps us understand the role of grains, grain boundaries, and other interfaces separately on the electrical conduction mechanism upon application of small amplitude ac signal[46]. This analysis also gives an insight into dielectric relaxation phenomena in ceramic [47,48]. The variation of imaginary part ($Z''$) of complex impedance ($Z$) with frequency range 40 Hz- 5 MHz at different temperatures (80 K-300 K) are shown in Figs 9a and 9c for CuO and $Cu_{0.95}Fe_{0.05}O$, respectively. Strong relaxation peaks have been observed for both samples. The intensity of the peaks gradually reduces with increasing temperature as the impedance also decreases with a rise in temperature. The peaks are shifted towards high frequencies with increasing temperature, indicating a thermally activated non-Debye-like relaxation process[49]. The possible reason for the shifting is the increment in the hopping rate of localized charge carriers at high temperatures [50]. AC conduction through grain boundaries and grain cause two relaxation peaks at a comparatively lower frequency and others at a higher frequency, respectively[51,52]. Only one relaxation peak has been originated at a lower frequency at each temperature in the case of CuO. At room temperature (300 K), it shows the relaxation peak at 902 Hz, and the peak shifted towards the lower frequency with decreasing temperature. No peak is observed after 160 K (at 50 Hz) towards lower temperature in our measured frequency range. As mentioned above, this relaxation peak can be attributed to electrical conduction through grain boundaries. No peak for grain conduction relaxation has been found. In the SEM study (Figure 4a) earlier, we observed small grains in CuO; possibly for this, more grain boundaries are dominating the conduction due to grains.

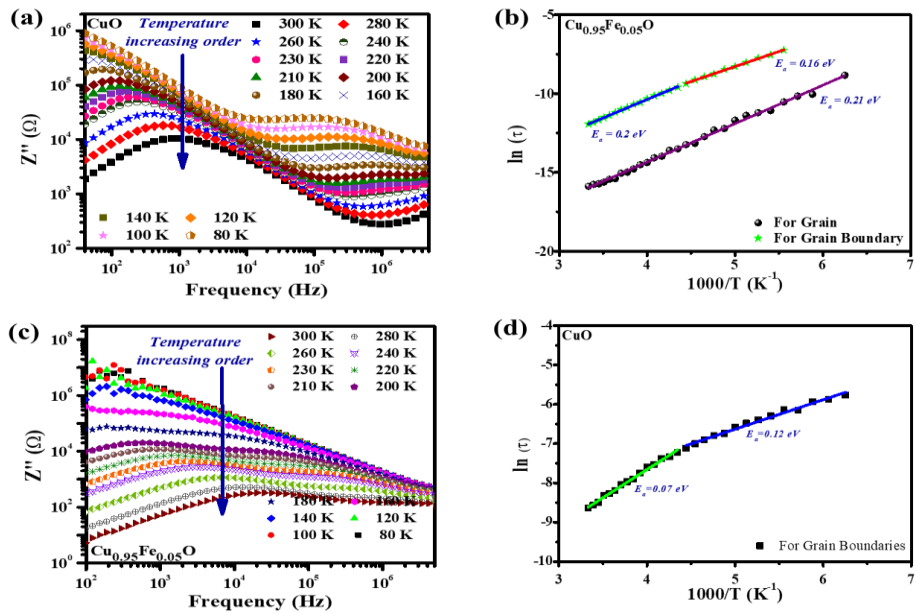

**Figure 9.** Frequency-dependent $Z''$ studies with temperature of (a) CuO and (c) $Cu_{0.95}Fe_{0.05}O$, the Arrhenius plots for grain boundaries and grains of (b) CuO and (d) $Cu_{0.95}Fe_{0.05}O$.

Whereas, Fe doped CuO sample has shown two relaxation peaks: one clear peak at a lower frequency and other hump type peak at a higher frequency at each temperature from 300 K up to 180 K. These peaks also showed the exact nature of shifting towards lower frequency with decreasing temperature, which indicates non-Debye like relaxation process[49]. At 300 K, the clear relaxation and hump type peaks are at 25.5 kHz and 1.1 MHz, respectively. After 180 K and 160 K, towards lower temperature, there is no peak in our measured frequency range for grain boundaries and grains, respectively. Clear and hump are relaxation peaks for electrical conduction through grain boundaries and grains, respectively. The relaxation times ($\tau$) were determined from the relation $\omega_m\tau=1$, where $\omega_m=2\pi f_m$, $f_m$ is the corresponding relaxation frequency. Activation energies for grain and grain boundaries were evaluated from the Arrhenius plot of $\tau$ vs. the inverse of temperature following the given equation:

$$\tau = \tau_0 e^{\left(\frac{-E_a}{k_B T}\right)} \tag{2}$$

where $E_a$ is the activation energy, $\tau_0$ is the pre-exponential factor, $k_B$ is the Boltzmann constant, T is the absolute temperature. The Arrhenius plots for grain boundaries and grains of CuO and $Cu_{0.95}Fe_{0.05}O$ are shown in Figs. 9b and 9d, respectively. The calculated activation energies are shown in Tables 1 for CuO and $Cu_{0.95}Fe_{0.05}O$. We observed two activation energies for the grain boundary conduction process for two temperature ranges.

**Table 1.** Activation energies of CuO and $Cu_{0.95}Fe_{0.05}O$

| CuO | $Cu_{0.95}Fe_{0.05}O$ | |
|---|---|---|
| $E_a$ for Grain boundaries in eV (T range in K ) | $E_a$ for Grain boundaries in eV (T range in K ) | $E_a$ for Grain in eV (T range in K) |
| 0.06 ± 0.002 (160-225) | 0.16 ± 0.002 (180-225) | 0.21 ± 0.002 (160-300) |
| 0.12 ± 0.003 (230-300) | 0.2 ± 0.008 (230-300) | - |

Further analysis of the Cole-Cole plot supports the conclusion arrived from the existence of relaxation peaks and the vital information about the grains and grains boundaries. The contribution of bulk grain, grain boundaries, and partial electrodes in charge transport of macroscopic polycrystalline samples can be modeled with meshes of a resistance R parallel to a capacitor C each [53,54].

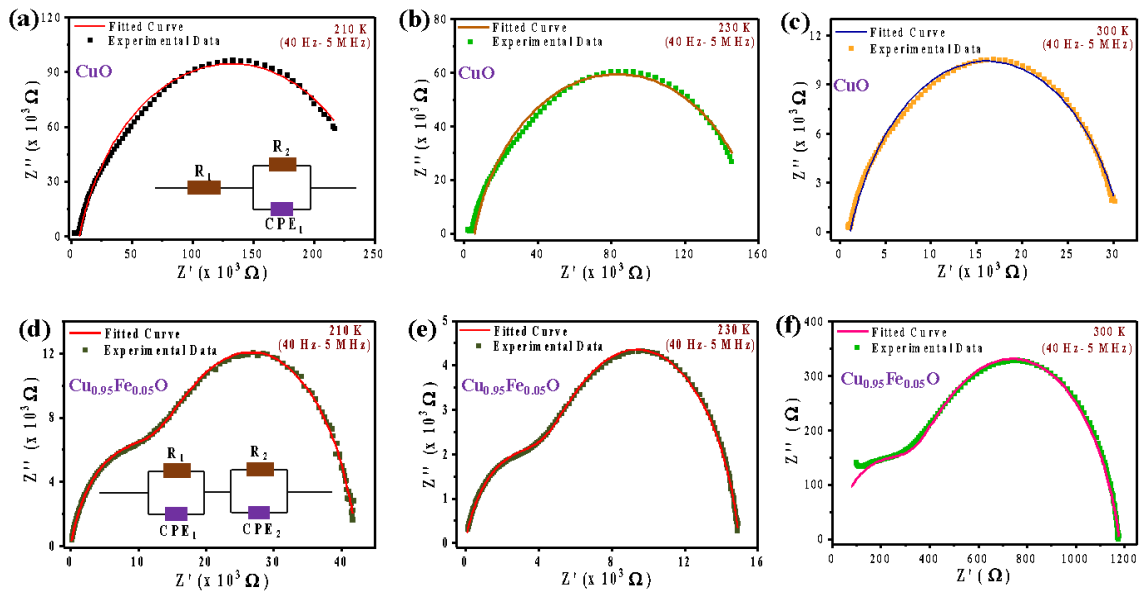

**Figure 10.** (a), (b), (c) Cole-Cole experimental data and fitted data plots of CuO at temperatures 210 K, 230 K and 300 K, respectively. (d), (e), (f) Cole-Cole experimental data and fitted data plots of $Cu_{0.95}Fe_{0.05}O$ at temperatures 210 K, 230 K and 300 K, respectively.

The Cole-Cole plots and its fitting through appropriate equivalent circuit for CuO at temperature 300 K and two transition temperatures 210 K and 230 K have been exhibited in Figs. 10(a-c). These plots showed only one depressed circular arc, instead of a good semicircle. This may be due to inhomogeneous function of grains and grain boundaries as a capacitor. Though, we have tried to follow the conventional fitting R-C element (resistance parallel with capacitor) but failed to fit properly. Therefore, we tried with a new fitting circuit replacing capacitor with a constant phase element (CPE). Appearance of CPE helped to understand the more real-system of solid-state ion conduction for ac signal.

**Table 2.** Fitted parameters of an equivalent circuit element of CuO

| Temperature in K | $R_1$ (% error) in $\Omega$ | $R_2$ (% error) in $\Omega$ | $Q_1$ (% error) | $n_1$ (% error) |
|---|---|---|---|---|
| 210 | 7010 (1.22) | $2.50 \times 10^5$ (0.88) | $1.46 \times 10^{-8}$ (1.1) | 0.82 (0.18) |
| 230 | 5573 (0.95) | $1.55 \times 10^5$ (0.87) | $1.52 \times 10^{-8}$ (1.3) | 0.83 (0.20) |
| 300 | 1205 (0.74) | 29807 (0.36) | $3.75 \times 10^{-8}$ (0.79) | 0.78 (0.11) |

The impedance (Z) of the circuit element i.e an ohmic resistance (R) in parallel with CPE (R || CPE) is given by [55]

$$Z = \left(\frac{1}{R} + \frac{1}{Z_{CPE}}\right)^{-1} \tag{3}$$

where, $Z = Z' + iZ''$, $Z_{CPE}$ is the impedance of CPE and $\frac{1}{Z_{CPE}} = Q(i\omega)^n$, $i = \sqrt{-1}$

$$Z' = \frac{R\left(1 + RQ\omega^n \coscos\left(\frac{n\pi}{2}\right)\right)}{1 + 2RQ\omega^n \coscos\left(\frac{n\pi}{2}\right) + (RQ\omega^n)^2} \tag{4}$$

$$Z'' = -\frac{R^2 Q\omega^n \sinsin\left(\frac{n\pi}{2}\right)}{1 + 2RQ\omega^n \coscos\left(\frac{n\pi}{2}\right) + (RQ\omega^n)^2} \tag{5}$$

The exponent n, may be varied between 0 to 1. When n=0 and 1, the CPE element is an excellent resistance and capacitor, respectively[55]. The Cole-Cole plots of CuO were perfectly fitted with the model shown in the inset of Fig. 10a. The resistance $R_1$ may be due to the influence of electrodes. The parameters determined from the fitting have been shown in Table 2. More interestingly, we have observed two depressed circular arcs in the case of Fe doped CuO, as expressed in Figs. 10(d-f). The minor arc towards high frequency is due to grains' contribution, and the large arc is for the contribution of grain boundaries. We try with two R|| CPE elements for grain boundaries and grains, and the theoretically generated data are matched

well with the experimental data, as shown in Figures 10(d-f). These observations of one and two arcs are strongly correlated with one and two relaxation peaks in the variation of Z'' with frequencies in the case of CuO and $Cu_{0.95}Fe_{0.05}O$, respectively.

Table 3. Fitted parameters of equivalent circuit element of $Cu_{0.95}Fe_{0.05}O$

| Temperature in K | $R_1$ (% error) in Ω | $R_2$ (% error) in Ω | $Q_1$ (% error) | $Q_2$ (% error) | $n_1$ (% error) | $n_2$ (% error) |
|---|---|---|---|---|---|---|
| 210 | 13589 (0.31) | 28721 (0.23) | $7.33 \times 10^{-9}$ (1.13) | $2.31 \times 10^{-8}$ (0.52) | 0.75 (0.12) | 0.85 (0.06) |
| 230 | 10372 (0.18) | 4636.5 (0.28) | $2.90 \times 10^{-8}$ (0.49) | $1.18 \times 10^{-8}$ (1.13) | 0.85 (0.06) | 0.74 (0.11) |
| 300 | 354.73 (0.59) | 823.51 (0.33) | $2.29 \times 10^{-8}$ (3.09) | $5.95 \times 10^{-8}$ (1.18) | 0.73 (0.26) | 0.83 (0.12) |

Table 3 exhibits the fitted parameters for $Cu_{0.95}Fe_{0.05}O$. It is noteworthy to mention that the determined values of $n_1$ and $n_2$ are close to 1 for all two samples, which denotes nearly the capacitive behavior of the CPE element. The heterogeneity of electrode surface leads to the cause of CPE element instead of capacitor [55].

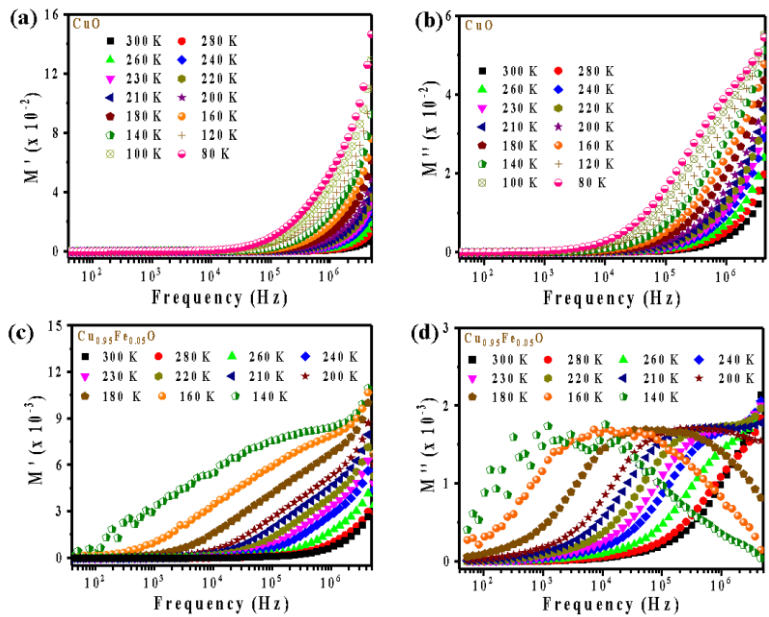

**Figure 11.** (a) and (b) Variation of M' and M'' with frequency at different temperature of CuO, respectively. (c) and (d) Variation of M' and M'' with frequency at different temperature of $Cu_{0.95}Fe_{0.05}O$, respectively.

Electric modulus analysis was performed to acquire more information on polarization types and electrical transport. The magnitude of the real part of this modulus (M') of CuO was approaching zero in the lower frequencies indicating the influence of interface was tending to be eliminated, as depicted in Figure 11a [56]. Then these values at each temperature have increased to their maximum magnitudes. This behavior can be attributed to conduction for short-range charge mobility [56,57]. In the case of M'' of CuO, we do not find any relaxation peak in Figure 11b in our measured frequency range. After iron doping, nature has been changed drastically. We have noticed a saturation tendency in M' value in the iron-doped CuO sample after reaching a maximum value. This saturation trend is clear at a comparatively lower temperature, as shown in Fig. 11c. As discussed for pure CuO, short-range mobility is responsible for this kind of behavior. Short-range mobility indicates the lack of retarding force on the charge carriers in the externally given a.c. Electric field during conduction in hopping regime is a mode of charge transfer. The contribution of electrons dominates the hole due to their high mobility than a hole in the conduction process. We have observed the relaxation peak in the variation of M'' with frequencies at all temperatures for $Cu_{0.95}Fe_{0.05}O$, as expressed in Fig. 10d. Before reaching the maximum relaxation frequency, the charge carriers can roam freely over long distances indicating conduction in the band regime. After this maximum frequency, the carriers have been confined in a potential well. So, it can be mentioned that the peak frequency denotes the transition from long-range to short-range mobility[58]. We may conclude the existence of a temperature-dependent hopping mechanism seeing this tendency. Due to iron doping, the charge transfer can happen quickly for the presence of excess ions. This might be a reason for saturation and relaxation peak of M' and M'', respectively, in our measured frequency range for Fe doped CuO. Whereas in pure CuO, we don't see this kind nature in M' and M'' except a sharp increment towards higher frequencies in our frequency range, beyond this range, one should have observed the same $Cu_{0.95}Fe_{0.05}O$.

### 3.7. Functional Group Analysis

In order to understand the chemical and structural nature of the synthesized pristine CuO and Fe doped CuO, Fourier Infra-red spectroscopy (FTIR) spectra analysis was carried out. Figure 12, represents the FTIR spectrum recorded for both samples in the range of 450 to 4,000 cm$^{-1}$. Vibration bands below 1000 cm$^{-1}$ is attributed to interatomic vibrations from the metal oxide [59]. Characteristic vibration bands located at ~ 534 cm$^{-1}$ and ~ 601 cm$^{-1}$ correspond to the vibrations arising from metal oxide stretching of monoclinic CuO. Higher frequency vibrational band at ~ 601 cm$^{-1}$ may be attributed to Cu-O stretching along [-101] direction, whereas ~ 534 cm$^{-1}$ band corresponds to the Cu-O stretching occurring along [101] direction [59]. Additionally, the presence of the Cu$_2$O phase (vibrational band located at ~ 610 cm$^{-1}$) is also overruled due to the absence of any vibrational band in the range of ~ 605-660 cm$^{-1}$ consistent with XRD results. Interestingly, Fe doping results in shifting of peak to the higher wavenumber and becomes broader, indicating that Fe ion is incorporated in the CuO lattice and leads to distortion of the crystal lattice.

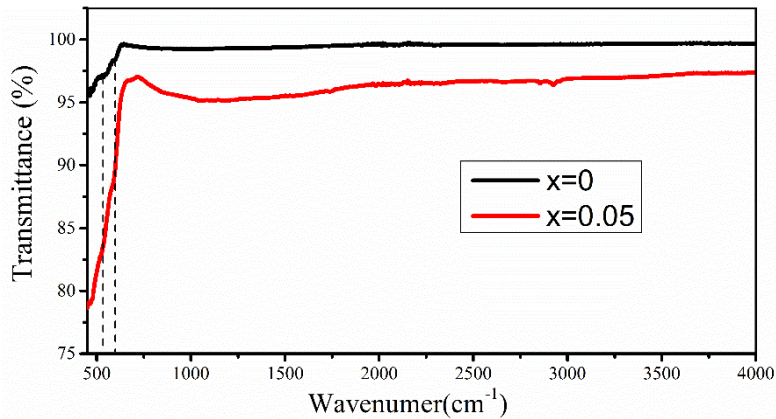

**Figure 12.** FTIR spectra of (a) CuO and (b) Cu$_{0.95}$Fe$_{0.05}$O ceramics at room temperature.

## 4. Conclusion:

In summary, we synthesized copper (II) oxide and trivalent iron-doped CuO through the solid-state method. Fe$^{3+}$ has promoted the paramagnetic phase in CuO. The two magnetic transition temperatures observed here indicate the transformation from paramagnetic phase to incommensurate or asymmetrical antiferromagnetic (AF) near highest Neel temperature ($T_{N1}$) ~230 K and the order transformation of AF phase to commensurate AF phase at around the Neel temperature ($T_{N2}$) ~210 K. The anomaly in the first-order derivative of dielectric constant near two $T_N$ has been perceived, which is due to the strong correlation between magnetic and dielectric properties. This supports the existence of spin-polaron coupling near Neel temperatures. The temperature-dependent Z'' plots with the frequency range and Cole-Cole plots suggest the strong effect of grain and grain boundaries in conduction. The role grain and grain boundaries are justified through an R|| CPE circuit element with the help of theoretical analysis. Electric modulus studies reveal the easy charge transfer for electrical conduction for iron-doped samples. The high dielectric constant with low loss and high magnetic susceptibility in Cu$_{0.95}$Fe$_{0.05}$O would reveal its future as spintronics and multifunctional material.